\documentclass[prl,twocolumn,superscriptaddress,showpacs]{revtex4-1}
\usepackage{graphics}
\usepackage{color}
\usepackage{amsmath}
\usepackage{amssymb}
\usepackage{hyperref}
\usepackage{graphicx}

\begin{document}

\title{Compass Impurity Model of Tb Substitution in Sr$_{2}$IrO$_{4}$}

\author{Long Zhang}
\affiliation{International Center for Quantum Materials, School of Physics, Peking University, Beijing, 100871, China}

\author{Fa Wang}
\affiliation{International Center for Quantum Materials, School of Physics, Peking University, Beijing, 100871, China}
\affiliation{Collaborative Innovation Center of Quantum Matter, Beijing, 100871, China}

\author{Dung-Hai Lee}
\affiliation{Department of Physics, University of California, Berkeley, CA 94720, USA}
\affiliation{Materials Sciences Division, Lawrence Berkeley National Laboratory, Berkeley, CA 94720, USA}


\begin{abstract}

We show that upon Tb substitution the interaction between the magnetic moments on the impurity Tb$^{4+}$ ion and its surrounding Ir$^{4+}$ ions is described by a ``compass'' model, i.e., Ising-like interaction favoring the magnetic moments across each bond to align along the bond direction. Such interaction nucleates quenched magnetic vortices near the impurities and drives a reentrant transition out of the antiferromagnetic ordered phase at low temperatures hence quickly suppresses the N\'eel temperature consistent with the experiment [Phys. Rev. B \textbf{92}, 214411 (2015)]. As a by-product, we propose that the compass model can be realized in ordered double perovskites composed of the spin-orbital-coupled $d^{5}$ ions and the half-closed-shell $f^{7}$ ions.

\end{abstract}
\date{\today}

\maketitle

\emph{Introduction.---}The layered iridate compound Sr$_{2}$IrO$_{4}$ has attracted much attention recently partly due to its close resemblance to the cuprate superconductors \cite{Kim2008a, Jackeli2009, Watanabe2010, Wang2011e}. It is described by a one-band (pseudo)spin-$1/2$ Hubbard model like the isostructural cuprate parent compound La$_{2}$CuO$_{4}$ \cite{Wang2011e}. Therefore, many interesting phenomena common to the cuprates are also found in Sr$_{2}$IrO$_{4}$. For example antiferromagnetic (AF) order exists in the parent compound \cite{Gum2009, Dhital2013, Ye2013c}, and upon doping Fermi arcs are seen in angle-resolved photoemission spectroscopy (ARPES) \cite{Kim2014a}. Moreover both scanning tunneling spectroscopy \cite{Yan2015a} and ARPES \cite{Kim2015c} suggest a low temperature nodal gap. Whether the latter is due to superconductivity is currently actively investigated.

The microscopic origins of the effective one-band description in Sr$_{2}$IrO$_{4}$ and cuprates, however, are quite different. For example in the parent compound the half filled band in cuprates has mixed copper $d_{x^{2}-y^{2}}$ and oxygen $p_{x,y}$ characters \cite{Zhang1988a}. For most purposes spin-orbit coupling (SOC) is negligible. In contrast, the relevant band of Sr$_{2}$IrO$_{4}$ derives from an effective total angular momentum (pseudospin) $1/2$ spin orbit coupled crystal field orbital. Due to the narrow bandwidth even the relatively weak on-site Coulomb correlation can render it Mott insulating \cite{Kim2008a}. At low temperatures the magnetic moments associated with this band become AF long range ordered \cite{Gum2009, Dhital2013, Ye2013c}.

In a recent experiment \cite{Wang2015d} Tb$^{4+}$ impurities are substituted into Sr$_{2}$IrO$_{4}$ to replace the Ir$^{4+}$ ions. The N\'eel temperature is fully suppressed by less than $3\%$ Tb substitution. At low temperatures the hysteretic magnetic susceptibility and the linear-$T$ specific heat behaviors suggest the formation of a spin glass state. These phenomena are reminiscent of the insulating lightly hole-doped cuprates, in which the N\'eel order is also replaced by a spin glass state at low temperatures \cite{Keimer1992, Chou1993, Niedermayer1998, Coneri2010}. Theoretically it has been shown the doped holes create nonlocal dipolar distortions, {i.e.}, magnetic vortex-antivortex pairs around the holes, which quickly destroy the AF order \cite{Aharony1988, Glazman1989, Glazman1990, Cherepanov1999a, Shraiman1988, Kou2003c, Kou2003b, Mei2013}. However, we expect the microscopic mechanism for the Tb substituted Sr$_{2}$IrO$_{4}$ to be different because the isovalent Tb$^{4+}$ substitution does not introduce extra charge carriers in the IrO$_{2}$ plane.

\begin{figure}[b]
\centering
\includegraphics[width=0.48\textwidth]{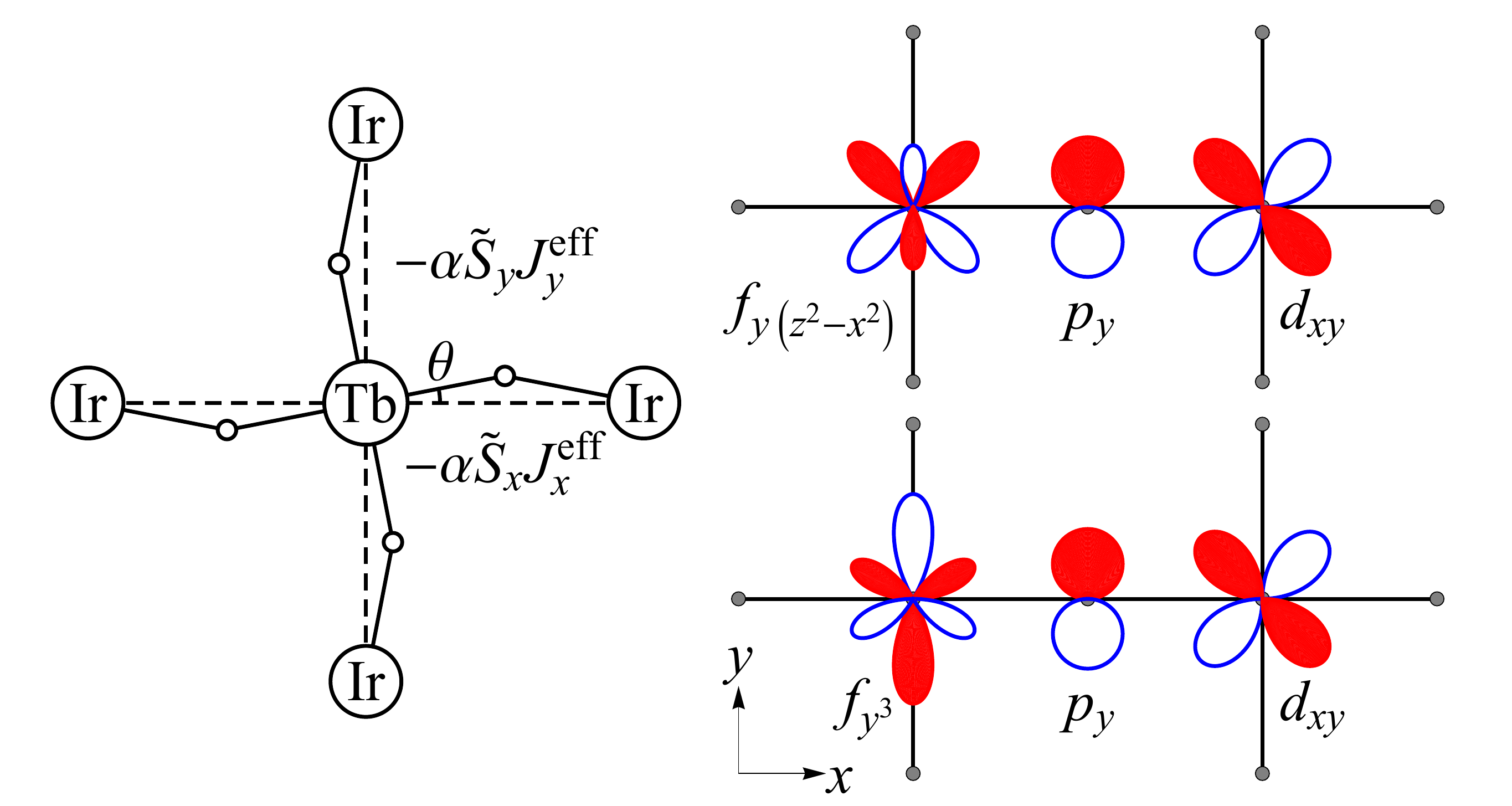}
\caption{Left: Configuration of the Tb$^{4+}$ impurity surrounded by four Ir$^{4+}$ ions. The lattice is slightly distorted, $\theta\simeq 11^{\circ}$. The magnetic interaction is Ising-like on each bond. Right: The bond geometries of the symmetry-allowed hopping processes $t_{1}$ (upper) and $t_{2}$ (lower). The signs of the wavefunctions are indicated by the filled (positive) and empty (negative) lobes.}
\label{fig:Tb}
\end{figure}

In this work, we first show that the magnetic interaction between the Tb$^{4+}$ impurity and its surrounding Ir$^{4+}$ ions is given to a good approximation by
\begin{equation}
\label{eq:hci}
H_{\mathrm{ci}} = -\alpha \sum_{i\in \mathrm{NN(Tb)}} \tilde{S}_{\gamma_{i}}J^{\mathrm{eff}}_{i,\gamma_{i}},
\end{equation}
in which the summation runs over the nearest neighboring Ir$^{4+}$ sites ($i=\pm\hat{x}, \pm\hat{y}$) of the Tb site (see Fig.~\ref{fig:Tb}, left panel). $\tilde{S}_{\gamma_{i}}$ is the spin operator of the Tb$^{4+}$ ion (slightly rotated due to the lattice distortion as will be discussed later) and $J^{\mathrm{eff}}_{i,\gamma_{i}}$ is the pseudospin operator of the Ir$^{4+}$ ion at site $i$. $\gamma_{i}=x$ ($y$) for $i=\pm\hat{x}$ ($\pm\hat{y}$). The Ising-like interaction on each bond favors the magnetic moments aligning along the bond direction like the compass model \cite{Nussinov2004, Nussinov2005, Doucot2005}, so we call it the \emph{compass impurity} model.  As we shall show such highly anisotropic magnetic interaction is rooted in the spin-orbital coupled nature of the Ir$^{4+}$ pseudospin-$1/2$ atomic levels. In previous studies such Ising-like magnetic interaction can arise only from edge-sharing IrO$_{6}$ octahedra \cite{Jackeli2009}. Our result opens a new route to strong exchange anisotropy in iridates with the corner-sharing IrO$_{6}$ octahedron structure.

This realization leads us to propose that the (uniform) compass model, which has topologically protected double degeneracy and may serve in quantum computation as qubits protected against decoherence \cite{Doucot2005}, can be realized in ordered double perovskites composed of spin-orbital-coupled $d^{5}$ ions (Ir$^{4+}$, Rh$^{4+}$, Ru$^{3+}$, etc.) and half-closed-shell $f^{7}$ ions (Tb$^{4+}$, Gd$^{3+}$, Eu$^{2+}$, etc.).

In an antiferromagnet with easy-plane anisotropy, {e.g.}, Sr$_{2}$IrO$_{4}$, the ``compass impurity'' induces a distortion of the AF order parameter which decays as $r^{-2}$ away from the impurity. In the dual Coulomb gas picture of the XY model the impurities induce quenched vortex quadrupoles. In the following we shall show the thermal vortices triggered by the impurity quadruple potential causes a low temperature reentrant transition to a disordered phase for arbitrarily small impurity concentration. Moreover, the N\'eel order is fully suppressed by only a few percent substitutions consistent with the experiment. Further experimental predictions shall also be discussed.

\emph{Compass impurity model.---}In the dilute impurity limit we first consider a single Tb$^{4+}$ ion embedded in the IrO$_{2}$ as shown in the left panel of Fig.~\ref{fig:Tb}. The Tb$^{4+}$ ion has electronic configuration $4f^{7}$. Because the $4f$ orbitals are very localized the crystal field effects is negligible and all $f$ orbitals are nearly degenerate. The Hund's rule coupling leads to a large spin $S=7/2$ and an orbital singlet state on each Tb$^{4+}$ ion.

The Tb-O-Ir bonds are slightly distorted due to the rotation of the IrO$_{6}$ and the TbO$_{6}$ octahedra around the $z$ axis. The rotation angle $\theta_{i}=\pm \theta$ ($\theta\simeq 11^{\circ}$) for $i\in A$ and $B$ sublattices respectively. Because the electric field perpendicular to the bond (which spoils the spin conservation in the hopping process) and the wavefunction overlap are only affected to order $\sin \theta$, as a good approximation we consider the bond to be straight and perform symmetry analysis in the usual (orbital, spin) basis.

In the undistorted case the Tb-O-Ir bond along $x$ direction has reflection symmetries along $y$ and $z$ axes, which allows the following nearest neighbor hopping parameters to be non-zero:  
\begin{equation}
\label{eq:t}
\begin{split}
t_{1}&= {}_{o}\langle f_{y(z^{2}-x^{2})}|H_{t}|d_{xy}\rangle_{\hat{x}},\quad t_{2}= {}_{o}\langle f_{y^{3}}|H_{t}|d_{xy}\rangle_{\hat{x}};\\
t'_{1}&= {}_{o}\langle f_{z(x^{2}-y^{2})}|H_{t}|d_{zx}\rangle_{\hat{x}},\quad t'_{2}= {}_{o}\langle f_{z^{3}}|H_{t}|d_{zx}\rangle_{\hat{x}}.
\end{split}
\end{equation}
Here $|d_{a}\rangle_{\hat{x}}$ and $|f_{b}\rangle_{o}$ denote the Ir$^{4+}$ orbitals at $\hat{x}$ and Tb$^{4+}$ orbitals at the origin respectively. We note that $t_{3}= {}_{o}\langle f_{xyz}|H_{t}|d_{yz}\rangle_{\hat{x}}$ is also allowed by symmetry, but it is much smaller than those in Eq. (\ref{eq:t}). The reasons are two fold: (1) the direct wavefunction overlap of the $d_{yz}$ and the $f_{xyz}$ orbitals is much smaller due to the $d_{yz}$ orbital orientation, (2) all possible oxygen $p$ orbitals mediated hopping processes are prohibited. Therefore we neglect the $t_{3}$ term in the rest of this work.

The $90^{\circ}$ rotation around the $x$ axis is also an approximate symmetry of the Tb-O-Ir bond if the electrons are well localized on the Ir and Tb ions. It relates the hopping parameters in Eq. (\ref{eq:t}) such that $t'_{1}\simeq t_{1}$ and $t'_{2}\simeq t_{2}$. This symmetry is well respected in the Ir-O-Ir bond of Sr$_{2}$IrO$_{4}$: the nearest-neighbor hopping parameters of the $d_{xy}$ and the $d_{zx}$ bands, which are also related by this rotation, are nearly equal: $t_{xy}=0.36~\mathrm{eV}$ and $t_{zx}=0.37~\mathrm{eV}$ \cite{Watanabe2010}. The Tb$^{4+}$ $4f$ orbitals are more localized, so this approximate symmetry should also be respected. Therefore, the nearest neighbor hopping between the Tb$^{4+}$ and the Ir$^{4+}$ ions along the $\hat{x}$ bond is described by
\begin{equation}
\label{eq:htx}
\begin{split}
H_{t,\hat{x}}=&\sum_{\sigma}\Big(t_{1}f_{y(z^{2}-x^{2}),\sigma}^{\dag}c_{\hat{x},xy,\sigma}+t_{2}f_{y^{3},\sigma}^{\dag}c_{\hat{x},xy,\sigma}\\
&+t_{1}f_{z(x^{2}-y^{2}),\sigma}^{\dag}c_{\hat{x},zx,\sigma}+t_{2}f_{z^{3},\sigma}^{\dag}c_{\hat{x},zx,\sigma}\Big)+\mathrm{H.c.}
\end{split}
\end{equation}
Here $\sigma$ is the spin component along the $z$ direction.

We then project Eq. (\ref{eq:htx}) onto the Ir$^{4+}$ pseudospin-$1/2$ atomic levels with the following replacement \cite{Wang2011e}: $c_{j,xy,\sigma}^{\dag}\rightarrow -i\sigma\sqrt{1/3}e^{i\theta_{j}\sigma/2}d_{j,\sigma}^{\dag}$ and $c_{j,zx,\sigma}^{\dag}\rightarrow \sigma\sqrt{1/3}e^{i\theta_{j}\sigma/2}d_{j,-\sigma}^{\dag}$ and find
\begin{equation}
\label{eq:htxp}
\begin{split}
H_{t,\hat{x}}=& \frac{1}{\sqrt{3}}\sum_{\sigma}\Big[\big(t_{1}f_{y(z^{2}-x^{2}),\sigma}^{\dag} +t_{2}f_{y^{3},\sigma}^{\dag} \big)i\sigma e^{i\sigma\theta /2}d_{\hat{x},\sigma}\\
& + \big(t_{1}f_{z(x^{2}-y^{2}),\sigma}^{\dag} +t_{2}f_{z^{3},\sigma}^{\dag}\big)\sigma e^{i\sigma\theta /2}d_{\hat{x},-\sigma}\Big] +\mathrm{H.c.}
\end{split}
\end{equation}
In the second term the \emph{pseudospin} is not conserved in the hopping process.  The reason is that while the spin is conserved [see Eq. (\ref{eq:htx})] the spin of the $d_{zx}$ component in the pseudospin-$1/2$ states is antiparallel to the pseudospin due to the SOC. This leads to the anisotropic magnetic interaction in Eq. (\ref{eq:hci}) as we shall see below. 

Taking into account the onsite Coulomb repulsion on the Ir$^{4+}$ and the Tb$^{4+}$ ions we derive the second order perturbation Hamiltonian and project it onto the $S=7/2$ subspace of the Tb$^{4+}$ ion. The effective magnetic interaction on the $\hat{x}$ bond is found to be $H_{\hat{x}}=-\alpha \tilde{S}_{x} J^{\mathrm{eff}}_{x,\hat{x}}$, in which $\tilde{S}_{x}=e^{i\theta S_{z}}S_{x}e^{-i\theta S_{z}}$ is the slightly rotated spin operator of the Tb$^{4+}$ ion. The interaction strength $\alpha=\frac{4}{21}(U_{d}^{-1}+U_{f}^{-1})(t_{1}^{2}+t_{2}^{2})$, in which $U_{d}$ [$U_{f}$] is the energy difference between the $(d^{6}, f^{6})$ [$(d^{4}, f^{8})$] and the $(d^{5}, f^{7})$ electron configurations due to the onsite Coulomb repulsion. The magnetic interaction on the Tb-O-Ir bond along the $y$ direction is derived in the same fashion, $H_{\hat{y}}=-\alpha \tilde{S}_{y} J^{\mathrm{eff}}_{y,\hat{y}}$ with $\tilde{S}_{y}=e^{i\theta S_{z}}S_{y}e^{-i\theta S_{z}}$. Combining $H_{\hat{x}}$ and $H_{\hat{y}}$ gives the compass impurity model, Eq.~(\ref{eq:hci}). The Hund's rule coupling on the Ir$^{4+}$ gives an AF Heisenberg-type correction, which is smaller than the compass-type interaction by one order of magnitude.

If Ir$^{4+}$ (or other ions with $d^{5}$ electronic configuration and strong SOC) and Tb$^{4+}$ (or other $f^{7}$ ions) form ordered double perovskites with the chemical formula A$_{4}$BB$'$O$_{8}$ (layered quasi-two dimensional structure) or A$_{2}$BB$'$O$_{6}$ (three dimensional structure), in which $\mathrm{B}=d^{5}$ ions and $\mathrm{B}'=f^{7}$ ions occupy different sublattices with unequal (pseudo)spin sizes in the 2D square or 3D cubic lattice, the effective magnetic interaction is given by the compass model,
\begin{equation}
H_{\mathrm{c}}=-\alpha \sum_{\langle ij\rangle}S_{\gamma_{ij},i}J^{\mathrm{eff}}_{\gamma_{ij},j},
\end{equation}
in which $\gamma_{ij} = x, y$ (and $z$ in the 3D case) for the bond $\langle ij\rangle$ along the $x, y$ (and $z$) directions respectively.

\emph{Impurity-induced quadrupolar distortion.---}The magnetic interaction of the pure Sr$_{2}$IrO$_{4}$ compound has an easy-plane anisotropy induced by the Hund's rule coupling on the Ir$^{4+}$ ions \cite{Jackeli2009, Watanabe2010}, so the magnetic moments form inplane AF order at low temperature \cite{Gum2009, Dhital2013, Ye2013c}. With Tb substitution such an anisotropy is strengthened by the compass-type interaction. For example, the magnetic moments of Heisenberg model with a single compass impurity align in the $xy$ plane in its classical ground state, which is obtained numerically. In experiments \cite{Wang2015d}, the uniform susceptibility along $c$ axis is larger than the inplane susceptibility at $x=0.03$, which also shows that the inplane AF correlation is stronger. Therefore, we shall study the impact of the compass impurities on the antiferromagnetic XY model.

\begin{figure}
\centering
\includegraphics[width=0.48\textwidth]{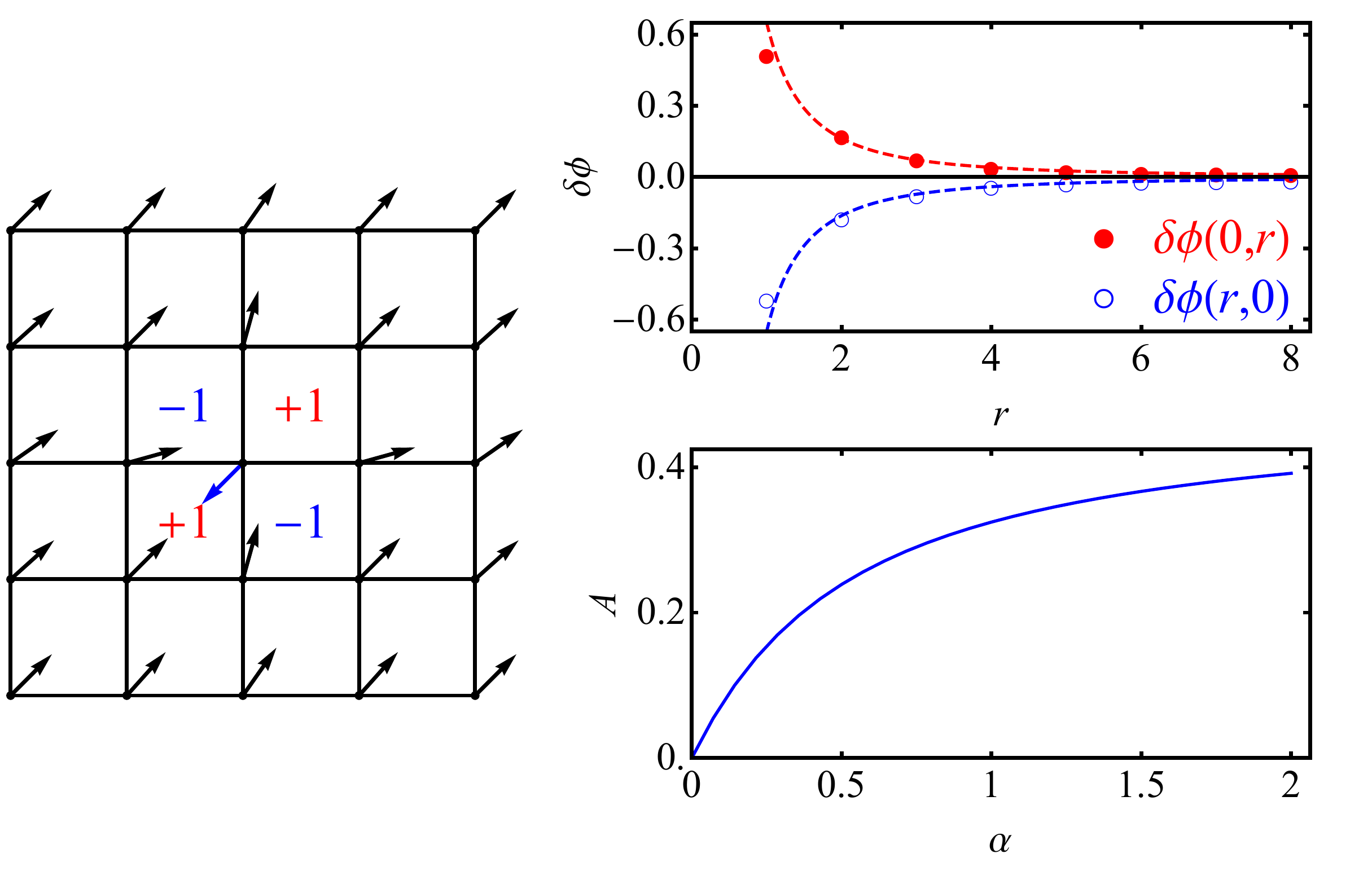}
\caption{Left: Staggered AF moment orientations in the classical ground state of Eq.~(\ref{eq:hciXY}) obtained by numerical minimization on a $17\times 17$-site lattice (only the central part is shown) for $J^{\mathrm{eff}}=1/2$, $S=7/2$, and $\alpha=1$. Vortices ($+1$) and antivortices ($-1$) appear in the plaquettes adjacent to the impurity site in the center. Upper right: Long range decay of AF moment distortion $\delta\phi(\mathbf{r})$ away from the impurity site. Dashed curves are fitting with Eq.~(\ref{eq:r2}). Lower right: Fitted quadrupole strength $A$ versus compass impurity strength $\alpha$.}
\label{fig:Quadrupole}
\end{figure}

The compass impurity induces local frustration to the AF order. We numerically calculate the classical ground state of a single compass impurity embedded in the XY antiferromagnet described by
\begin{equation}
\label{eq:hciXY}
H_{\text{ci-XY}}=\sum_{\langle ij\rangle'}\big(J^{\mathrm{eff}}_{x,i}J^{\mathrm{eff}}_{x,j}+J^{\mathrm{eff}}_{y,i}J^{\mathrm{eff}}_{y,j}\big)+H_{\mathrm{ci}},
\end{equation}
in which $\langle ij\rangle'$ excludes the bonds connected to the impurity site. The result is shown in Fig.~\ref{fig:Quadrupole}, left panel, where the arrows indicate the direction of the staggered magnetic moments.

Due to the {\it ferromagnetic} Ising-like interaction around the impurity, the AF moment on the impurity site lies antiparallel to the total AF moment of the system. The nearby AF moments are also distorted to gain the anisotropic interaction energy. This creates vortices and antivortices in the plaquettes adjacent to the impurity, {i.e.}, the AF moment orientation changes by $\pm 2\pi$ as one encircles the plaquette as shown in Fig.~\ref{fig:Quadrupole}, left panel \footnote{The vorticity in a single plaquette is counted by adding up the phase differences along the bonds $\phi_{ij}=\phi_{i}-\phi_{j}$, in which $\phi_{ij}$ are restricted to the interval $(-\pi, \pi]$ \cite{Jelic2011}.}. Therefore, the compass impurity induces a vortex \emph{quadrupolar}  distortion of the AF order. The long range behavior of the AF moment orientation $\phi(\mathbf{r})$ is given by the solution of the following equation, 
\begin{equation}
\epsilon_{ij}\partial_i\partial_j\phi(\mathbf{r})=AQ_{ij}\partial_{i}\partial_{j}\delta(\mathbf{r}),
\label{eq:laplace}
\end{equation}
in which $Q_{ij}$ is a normalized ($\det Q=-1$) traceless symmetric tensor indicating the orientation of the quadrupole moment and $A$ is the quadrupole strength. 

The solution of Eq.~(\ref{eq:laplace}) is $\phi(\mathbf{r})=\bar{\phi}+\delta\phi(\mathbf{r})$ with the distortion $\delta\phi(\mathbf{r})$ given by
\begin{equation}
\delta\phi(\mathbf{r})=2A\frac{r_{i}Q_{ij}\epsilon_{jk} r_{k}}{r^{4}},
\label{eq:r2}
\end{equation}
which fits the numerical results perfectly as shown in Fig.~\ref{fig:Quadrupole}, upper right panel. Therefore, the impurity-induced distortion to the AF order extends nonlocally and decays as $r^{-2}$.

The quadrupole strength $A$ is obtained by fitting Eq.~(\ref{eq:r2}) for different compass impurity strength $\alpha$ (Fig.~\ref{fig:Quadrupole}, lower right panel). It is of $O(1)$ order and increases monotonically with $\alpha$.

\emph{Suppression of AF order.---}The classical XY model without impurities has the famous Kosterlitz-Thouless (KT) transition at finite temperature, which is driven by the unbinding of thermally activated vortex-antivortex pairs \cite{Kosterlitz1973}. In the presence of quenched vortex dipoles a reentrant transition to a disordered phase occurs at low temperatures \cite{Rubinstein1983}.

The compass impurities with quenched vortex \emph{quadrupoles} turn out to have a similar impact on the AF order. In the presence of many quadrupolar impurities the continuum Hamiltonian is given by
\begin{equation}
H_{\phi}=\rho_{s} \int d^{2}\mathbf{r}\Big( \frac{1}{2}(\partial _{i}\phi(\mathbf{r}))^{2} -\partial_{i}\phi(\mathbf{r})f_{i}(\mathbf{r})\Big),
\end{equation}
in which $\rho_{s}$ is the spin stiffness. The second term describes the interaction of the AF moment field with the quadrupolar impurity potential,
\begin{equation}
f_{i}(\mathbf{r})=A\sum_{l}Q_{ik}(l)\epsilon_{kj}\partial_{j}\delta(\mathbf{r}-\mathbf{r}^{\mathrm{q}}_{l}).
\end{equation}
In the dilute impurity limit, the quadrupole-quadrupole interaction decays as $r^{-4}$ and can be neglected. Therefore, both the position $\mathbf{r}^{\mathrm{q}}_{l}$ and the orientation $Q_{ij}(l)$ of each quadrupole are treated as quenched random variables without spatial correlation. Upon disorder average (d.a.) we have
\begin{equation}
\begin{split}
&[f(\mathbf{r})]_{\mathrm{d.a.}}=0,\\
&[f_{i}(\mathbf{r})f_{i'}(\mathbf{r}')]_{\mathrm{d.a.}}=-\frac{1}{2} x A^{2}\delta_{ii'} \partial^{2}\delta(\mathbf{r}-\mathbf{r}'),
\end{split}
\end{equation}
in which $x$ is the impurity concentration.

In the dual Coulomb gas picture vortices and antivortices are mapped to electric charges and the impurity vortex quadruples are mapped to electric quadrupoles. The Hamiltonian is given by
\begin{equation}
\begin{split}
H_{\mathrm{v}}=&-\pi\rho_{s}\sum_{l\neq l'}m_{l}m_{l'}\log\left|{\mathbf{r}^{\mathrm{v}}_{l}-\mathbf{r}^{\mathrm{v}}_{l'}}\over a \right| +E_{c}\sum_{l}m_{l}^{2}\\
&+\rho_{s}\sum_{l}m_{l} \int d^{2}\mathbf{r} f_{i}(\mathbf{r})\frac{(\mathbf{r}^{\mathrm{v}}_{l}-\mathbf{r})_{i}}{(\mathbf{r}^{\mathrm{v}}_{l}-\mathbf{r})^{2}},
\end{split}
\label{eq:Coulomb}
\end{equation}
in which $a$ is the short-distance cutoff and $m_{l}$ is the vorticity at $\mathbf{r}^{\mathrm{v}}_{l}$. $E_{c}$ is the vortex core energy. The last term is the Coulomb interaction between the vortices and the quenched quadrupolar impurities. 

Following the standard KT renormalization group (RG) procedure \cite{Jose1977, Chaikin1995principles} we define the reduced spin stiffness $K=\rho_{s}/k_{\mathrm{B}}T$ and find, from the dielectric function, the renormalized stiffness $K_{\mathrm{R}}$ as
\begin{equation}
\label{eq:kr}
K_{\mathrm{R}}=K+\pi^{2}K^{2}\int_{a}^{\infty}\frac{d^{2}\mathbf{r}}{a^{2}}\frac{r^{2}}{a^{2}}[\langle m(\mathbf{r})m(\mathbf{0})\rangle_{T}]_{\mathrm{d.a.}},
\end{equation}
in which  $m(\mathbf{r})=\sum_{l}m_{l}\delta(\mathbf{r}-\mathbf{r}^{\mathrm{v}}_{l})$ is the charge density of the Coulomb gas,  and $\langle\cdot\rangle_{T}$ is the thermal average. Define the (thermal) vortex fugacity $y=e^{-E_{c}/k_{\mathrm{B}}T}$ we find, to $O(y^{2})$, 
\begin{equation}
\label{eq:mm}
\langle m(\mathbf{r})m(\mathbf{0})\rangle_{T}=-2y^{2}(r/a)^{-2\pi K}\cosh I(\mathbf{r}),
\end{equation}
in which
\begin{equation}
I(\mathbf{r})=2\pi K\int d^{2}\mathbf{r}' f_{i}(\mathbf{r}')\partial'_{i}(G(\mathbf{r}'-\mathbf{r})-G(\mathbf{r}')),
\end{equation}
where $G(\mathbf{r})=(1/2\pi)\log(r/a)$.

The disorder average can be evaluated using the cumulant expansion \cite{Rubinstein1983},
\begin{equation}
\label{eq:Ida}
[\cosh I(\mathbf{r})]_{\mathrm{d.a.}} = e^{\frac{1}{2} [I(\mathbf{r})^{2}]_{\mathrm{d.a.}}}=e^{2\pi^{2}x A^{2}K^{2}\delta_{a}},
\end{equation}
in which $\delta_{a}$ is the short-range regularization constant for the $\delta$ function, which comes from the core of the quenched quadruples.

From Eqs. (\ref{eq:kr})--(\ref{eq:Ida}) we find that the renormalized stiffness $K_{\mathrm{R}}$ has exactly the same form as in the standard KT transition if $y$ is replaced by $y_{x}$,
\begin{equation}
\label{eq:ylambda}
y_{x}=y e^{\pi^{2}x A^{2} K^{2}\delta_{a}}.
\end{equation}
The RG equations are given by \cite{Jose1977, Chaikin1995principles}
\begin{equation}
\begin{split}
&{d\over dl}y_{x}=(2-\pi K)y_{x}, \\
&{d\over dl}K^{-1}= 4\pi^{3}y_{x}^{2}.
\end{split}
\end{equation}

\begin{figure}
\centering
\includegraphics[width=0.48\textwidth]{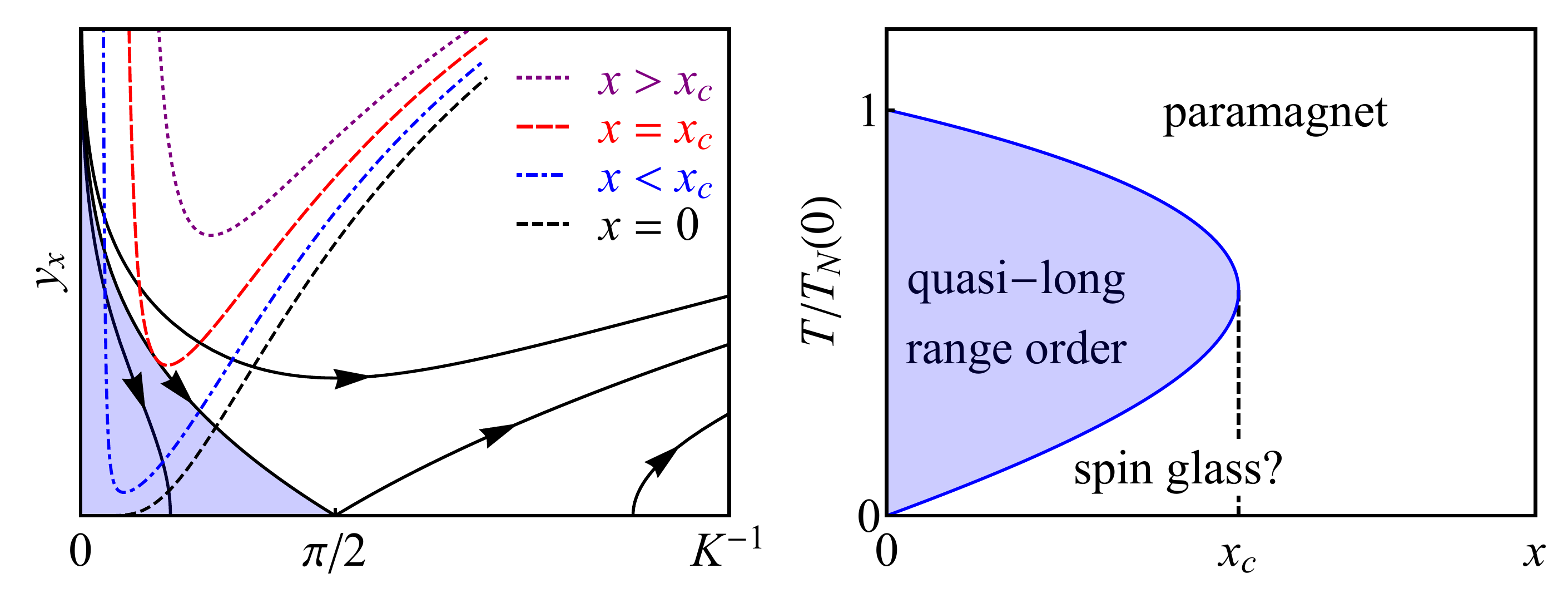}
\caption{Left: RG flow in the $y_{x}$-$K^{-1}$ parameter plane. The dashed curves illustrate the variation of $y_{x}$ with temperature for different impurity concentration $x$. Right: Schematic phase diagram of the XY model with quenched quadrupolar impurities. The shaded regions in both panels indicate the quasi-long range order phase.}
\label{fig:RG}
\end{figure}

The RG flow in the $y_{x}$-$K^{-1}$ parameter plane is shown in Fig.~\ref{fig:RG}, left panel. The shaded region, in which the reduced stiffness $K$ flows to a non-zero value, is the KT phase with quasi-long range order (QLRO). The region outside is disordered because $K$ flows to zero. In the absence of impurities $y_{x}$ reduces to the vortex fugacity $y$. Its variation with the temperature is drawn as the dashed black curve -- the pure system has QLRO at low temperature and becomes disordered at the KT transition, namely, the N\'eel temperature $T_{N}(0)$.

In the presence of quenched impurities the extra factor in $y_{x}$ represents the nucleation of vortices near the quadrupoles. Approaching zero temperature it diverges faster than how $y$ vanishes so the system is disordered at zero temperature for any impurity concentration $x$. The variation of $y_{x}$ with temperature is illustrated as the dashed curves for different impurity concentrations. There is a critical concentration $x_{c}$. Below $x_{c}$ the system shows QLRO at an intermediate temperature range and enters a reentrant disordered phase at low temperatures. Because this low-$T$ disordered regime is driven by the impurity potential we believe it can show the spin glass behavior seen in the experiment \cite{Wang2015d, Fischer1991spin}. Above $x_{c}$ the intermediate ordered phase vanishes and the N\'eel temperature $T_{N}$ abruptly drops to zero. This is schematically illustrated in the phase diagram (Fig.~\ref{fig:RG}, right panel).

The critical concentration $x_{c}$ is not universal. It depends on the vortex core energy $E_{c}$, the quadrupole strength $A$ and the inverse quadrupole core area $\delta_{a}$. If we take all these quantities to be of $O(1)$ order, $x_{c}$ is found to be only a few percent. For example, for $E_{c}=2$ and $A^{2}\delta_{a}=2$ we find $x_{c}=0.027$, which is consistent with the quick suppression of the N\'eel temperature in the experiment \cite{Wang2015d}.

\emph{Summary.---}To summarize, the magnetic interaction near Tb impurities in Sr$_{2}$IrO$_{4}$ is described by the planar compass impurity model. The strong in-plane anisotropy around the Tb site can be detected with nuclear magnetic resonance. The compass impurity induces a long range quadrupolar distortion to the antiferromagnetic order which drives a reentrant transition to a disordered phase at low temperature and quickly suppresses the N\'eel temperature. Motivated by this work we propose that the compass model can be realized in ordered double perovskites composed of spin-orbital-coupled $d^{5}$ ions and half-closed-shell $f^{7}$ ions.

\acknowledgements
L.Z. is grateful to J. C. Wang for helpful discussions. This work was supported by the National Key Basic Research Program of China (Grant No. 2014CB920902) and the National Natural Science Foundation of China (Grant No. 11374018). DHL is supported by DOE Office of Basic Energy Sciences, Division of Materials Science, under Material Theory program, DE-AC02-05CH11231.

\bibliography{library}

\end{document}